\documentclass[a4paper]{article}

\usepackage{kotex} 
\usepackage{cite}
\usepackage{INTERSPEECH2022}
\usepackage{multirow}
\usepackage{setspace}

\title{Adversarial Multi-Task Learning for Disentangling Timbre and Pitch in Singing Voice Synthesis}
\name{Tae-Woo Kim, Min-Su Kang, Gyeong-Hoon Lee}
\address{Speech AI Lab., AI Center, NCSOFT Corp., Republic of Korea}
\email{\{ktw0114, mskang, ghlee0304\}@ncsoft.com}

\begin{document}

\maketitle
\begin{abstract}
Recently, deep learning-based generative models have been introduced to generate singing voices. 
One approach is to predict the parametric vocoder features consisting of explicit speech parameters.
This approach has the advantage that the meaning of each feature is explicitly distinguished. 
Another approach is to predict mel-spectrograms for a neural vocoder. 
However, parametric vocoders have limitations of voice quality and the mel-spectrogram features are difficult to model because the timbre and pitch information are entangled. 
In this study, we propose a singing voice synthesis model with multi-task learning to use both approaches -- acoustic features for a parametric vocoder and mel-spectrograms for a neural vocoder. 
By using the parametric vocoder features as auxiliary features, the proposed model can efficiently disentangle and control the timbre and pitch components of the mel-spectrogram. 
Moreover, a generative adversarial network framework is applied to improve the quality of singing voices in a multi-singer model.
Experimental results demonstrate that our proposed model can generate more natural singing voices than the single-task models, while performing better than the conventional parametric vocoder-based model.

\end{abstract}

\section{Introduction}
Singing voice synthesis (SVS) is a generative model to synthesize singing voices according to musical scores and lyrics. 
Although the musical scores provide note pitch and duration information, it is difficult to model singing voices as they have a longer vowel duration and wider pitch range than speech signals in general. 
Moreover, singing voices have timbre, which is everything except the pitch or loudness\cite{svs}, and pitch information that can change according to the singer’s vocalizations and expressions.
The singer’s timbral characteristics depend on the voice source and vocal tract\cite{sundberg1990science}.
To model natural singing voices, SVS is modeled considering the interdependence between timbre and pitch. 

In recent years, the designs of most SVS models\cite{xiaoicesing, xiaoicesingmulti, hifisinger, atk, disentangle, nsinger, begansing, mlpsinger} have been inspired by the neural network-based architecture in text-to-speech (TTS) models\cite{fastspeech, fastspeech2, fastpitch, fastpitchformant, tacotron1, tacotron2, dctts}.
Recent Korean SVS systems\cite{atk, disentangle, nsinger} are able to generate mel-spectrograms directly by disentangling the timbre and pitch related spectra without external signal processing to generate realistic and natural singing voice.
Lee et al.\cite{atk} have proposed an adversarial trained end-to-end Korean SVS system that is based on Deep Convolutional TTS\cite{dctts} and includes a phonetic enhancement masking decoder that predicts the formant spectral mask. 
In \cite{disentangle}, the singer's identity is conditioned on each independent decoder of \cite{atk} to generate the timbre and singing style.
N-Singer\cite{nsinger}, a non-autoregressive SVS model, independently models the phoneme and pitch modules to generate accurately pronounced singing voices.
However, in these methods\cite{atk, disentangle, nsinger}, linguistic and note information is fed independently in each module to separate the timbre and pitch representations in an unsupervised manner.
Additionally, they have limitations in generating natural singing voices because timbre and pitch are assumed to be independent of each other.

In this study, we propose an adversarial multi-task learning-based SVS model to disentangle the timbre and pitch representations. 
The main task of the proposed method is to predict the mel-spectrogram, while the auxiliary task is to predict the features of timbre and pitch. 
The proposed model has been trained in two phases and adversarial trained using discriminators in both phases.
In the first phase, it is pre-trained on the auxiliary task.
This pre-training allows the two decoders to represent timbre and pitch, respectively.
In the second phase, it is jointly trained on the main and auxiliary tasks.
Thus, the disentangled mel-spectrograms for timbre and pitch are predicted by the respective decoders and integrated into the final mel-spectrogram.
In the experiments, the proposed model performs better than the single-task SVS models in terms of perceptual quality and naturalness.
Further, synthesizing the timbre and pitch features as predicted by the auxiliary task using the WORLD vocoder\cite{world} performs better than the conventional WORLD vocoder-based SVS model.

\section{Related works}
Multi-task learning (MTL) has been widely used in speech applications related to enhancement\cite{koizumi2020speech, eskimez2021human}, recognition\cite{pironkov2016multi} and synthesis\cite{wu2015deep, liu2020modeling}.
It is a learning paradigm that improves the performance of generalizations by sharing related knowledge from jointly learning multiple tasks\cite{MTL}. 
The model proposed in this study applies MTL to the SVS task to jointly learn two tasks -- WORLD vocoder and neural vocoder feature prediction.

Generative adversarial networks (GAN) are a powerful method to solve over-smoothing problems and generate high-resolution images\cite{gan}.
In speech synthesis, Yang et al.\cite{ganspeech} improved speech quality by jointly training a generator and a conditional discriminator using the speaker embeddings in multi-speaker models.
We adopt the conditional GAN framework using singer embeddings for the multi-singer SVS model, as in \cite{ganspeech}. 

\begin{figure*}[t]
  \centering
  \includegraphics[width=\linewidth]{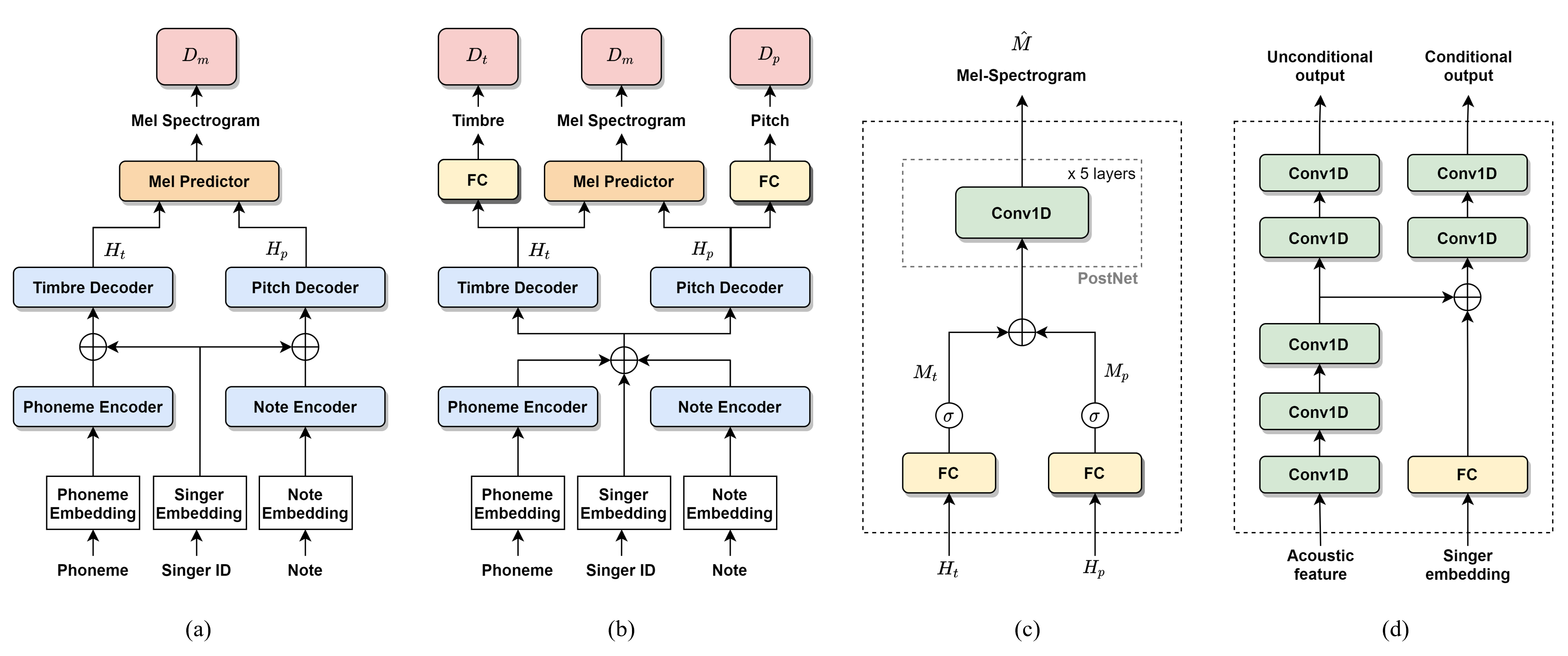}
  \caption{Architecture overview. (a) is a single-task SVS model, (b) is the proposed multi-task SVS model, (c) is a mel-predictor, and (d) is a discriminator.}
  \label{fig:overview}
\end{figure*}

\section{Proposed methods}
As depicted in Figures~\ref{fig:overview}(a) and (b), the generators of the single-task SVS model and proposed multi-task SVS model consist of two encoders, two decoders, and a mel-predictor.
The single-task model is based on the architecture of N-Singer\cite{nsinger} wherein phonemes and note pitches are independently modeled. 
Unlike the single-task model, the proposed model's encoder outputs are integrated such that the timbre and pitch decoders are guided through the auxiliary tasks, thereby enabling them to represent each characteristic.
Then, predicted timbre and pitch representations are integrated by the mel-predictor to predict the final mel-spectrogram.
Thus, the mel-spectrogram and auxiliary features are adversarial trained with discriminators.

\subsection{Embeddings}
As with previous Korean SVS systems\cite{atk, disentangle, nsinger, begansing, mlpsinger}, we utilize the advantage that one syllable in the lyrics matches one note of musical instrument digital interface (MIDI). First, we decompose Korean syllables into a phoneme sequence using the Korean grapheme-to-phoneme algorithm. 
Second, we get a frame-level phoneme sequence $T\in\mathbb{R}^{1 \times L}$ by assigning phoneme to frame corresponding to each phoneme using a MIDI note sequence $P\in\mathbb{R}^{1 \times L}$ that includes starting time, duration, and pitch, where $L$ is the length of the acoustic feature. 
Third, because each Korean syllable consists of an onset, a nucleus, and an optional coda in general, we assign the onset and coda to the first and last three frames, respectively, while assigning the remaining frames to the nucleus within the interval corresponding to each syllable in the phoneme sequence. 
The input sequences are embedded into $D$-dimensional dense vectors $E_{T}\in\mathbb{R}^{D \times L}$ and $E_{P}\in\mathbb{R}^{D \times L}$ using learnable lookup tables, respectively. 
The embedded phoneme sequence $E_{T}$ and note sequence $E_{P}$ are passed on to the phoneme encoder and note encoder respectively, after which they are integrated with the singer embedding vector $E_{S}\in\mathbb{R}^{D \times 1}$. 
The phoneme and note encoders consist of stacked conformer blocks\cite{conformer}. 
The singer embedding vector $E_{S}$ is obtained using a learnable lookup table.

\subsection{Timbre and pitch decoders}
The two decoders in our model consist of stacked conformer blocks to predict the timbre and pitch representations.
As the two decoders cannot disentangle the individual representations of timbre and pitch in an unsupervised manner owing to the common input received from the encoders, an MTL approach is introduced.
With the auxiliary task of explicitly predicting timbre and pitch features, the two decoder networks are shared to learn the timbre and pitch representations, respectively.
First, the timbre representation $H_{T}$, which is the output of the shared timbre decoder, is passed through the fully connected (FC) layer and the mel-predictor.
The FC layer predicts the timbre features -- mel-generalized cepstrum (MGC), band aperiodicity (BAP), and voiced/unvoiced (V/UV) flags.
On the other side, the pitch representation $H_{P}$, which is the output of the shared pitch decoder, is passed to the other FC layer predicting log-scale F0 (logF0) and to the mel-predictor. During training, the ground truth V/UV flags are applied to train the logF0 of the voiced section. Thus, the loss for the pitch is calculated as 
\begin{equation}
    L_{pitch} = L_{1}(p, \hat{p} \odot v),
    \label{eq1}
\end{equation}
where $\odot$, $p$, $\hat{p}$, and $v$ are the element-wise multiplication, target logF0, predicted logF0, and ground-truth V/UV flags, respectively. 
Then, the loss for all the auxiliary features is as follows:
\begin{equation}
\begin{split}
    L_{aux} = {}& w_{m} \cdot L_{mgc}+ w_{b} \cdot L_{bap} \\ 
      {}& + w_{v} \cdot L_{vuv} + w_{p} \cdot L_{pitch},
    \label{eq2}
\end{split}
\end{equation}

where $L_{mgc}, L_{bap}$ and $L_{vuv}$ indicate the loss of the MGC, BAP, and V/UV, respectively.
$L_{vuv}$ uses the binary cross-entropy function, while the others use the L1 loss function.
$w_m$, $w_b$, $w_v$, and $w_p$ are the respective scalar weights corresponding to each loss. 

\subsection{Mel-predictor}
As shown in Figure~\ref{fig:overview}(c), the timbre and pitch representations that are output from the two decoders are used to predict the log-scale mel-spectrogram by the mel-predictor.
First, the two hidden representations pass through the two respective FC layers in Figure~\ref{fig:overview}(c) to reduce dimensionality, after which the sigmoid is applied and output as the visible and interpretable representations, $M_{T}$ and $M_{P}$.
In our preliminary experiments, we found that the two representations equal the spectral envelope and pitch harmonics, respectively.
According to the source filter theory\cite{sourcefilter}, the pitch harmonics are multiplied by the spectral envelope in the linear spectral domain. 
However, since this study deals with log-scale mel-spectrograms, multiplication is replaced by a summation operation.
Therefore, $M_{T}$ and $M_{P}$ are summed and passed through a convolutional neural network-based postnet\cite{tacotron2} to predict the final mel-spectrogram.
Since the source filter's modeling is applied to the mel-spectrogram domain, the postnet in the mel-predictor makes the predicted coarse mel-spectrogram close to the true mel-spectrogram $M$.
Consequently, the loss for the mel-spectrograms is represented as:
\begin{equation}
\begin{split}
    L_{mel} = L_{1}(M, M_T + M_P) + L_{1}(M, \hat{M})
    \label{eq3}
\end{split}
\end{equation}

\subsection{Discriminators}
To improve the perceptual quality of the generated singing voices, we adopt the conditional GAN framework proposed in \cite{ganspeech}.
As shown in Figure~\ref{fig:overview}(d), we use the joint conditional and unconditional discriminator of \cite{ganspeech} with the singer embeddings.
Unlike previous models\cite{ganspeech} that have only used the GAN framework to enhance the mel-spectrogram, we use three discriminators for the mel-spectrogram, MGC and logF0 to further disentangle the timbre and pitch representations.
We have experienced that training for the auxiliary features using reconstruction loss is not sufficient to disentangle the timbre and pitch representations. 
Moreover, since MGC and logF0 in the multi-singer model have high diversities based on prosodic attributes such as the singer's identity, purely reconstruction loss-based training would weaken the capabilities of MGC and logF0 prediction.
That would result in $H_t$ and $H_p$ having entangled attributes between the timbre and pitch due to the powerful mel-predictor, leading to a degradation of performance.
Thus, we use the least-squares loss function \cite{LSGAN} and the additional feature matching loss $L_{fm}$ for the adversarial training as follows:
\begin{align}\label{eq4}
    L_{dis} &=\frac{1}{|S_F|} \sum_{i\in S_F}\left[ \frac{1}{2}\mathbb{E}_{s}[D_i(\hat{x}_i)^{2}+D_i(\hat{x}_i, s)^{2}] \right. \nonumber\\
    {}{} &+ \left. \frac{1}{2} \mathbb{E}_{(x,s)}[(D_i(x_i)-1)^{2}+(D_i(x_i, s)-1)^{2}] \right], \\
    L_{adv} &= \frac{1}{ |S_F| } \sum_{i\in S_F}\mathbb{E}_{s} \left [ (D_i(\hat{x}_i)-1)^{2} + (D_i(\hat{x}_i, s)-1)^{2} \right ],\\
    L_{fm} &= \frac{1}{|S_F|} \sum_{i\in S_F}\mathbb{E}_{(x, s)} \left [\sum_{l=1}^{L}\frac{1}{N_l}|| D_{i}^{(l)}(x_i)-D_{i}^{(l)}(\hat{x}_i)||_1 \right],
\end{align}

where $x$, $\hat{x}$, and $s$ are the target feature, generated feature, and singer embedding, respectively. $S_F = \{m, t, p\}$ is the index set including the mel-spectrogram, MGC, and logF0, while $L$ is the total number of layers in each discriminator. Accordingly, the total loss function of the generator as follows:
\begin{equation}
    \begin{split}
        L_{total} = L_{mel} + L_{aux} + L_{adv} + \lambda_{fm} L_{fm}
        \label{eq7}
    \end{split}
\end{equation} 

During the first pre-training phase, we remove $L_{mel}$ and set $S_F = \{t, p\}$ to further disentangle the timbre and pitch representations. 
For the multi-task training phase, we train our model using $L_{total}$ as Equation (\ref{eq7}).

\section{Experiments}
\subsection{Experimental setups}
\subsubsection{Dataset}
We collected singing voice data of 50 Korean songs each from both singers (one male and one female) in the children's song style (CS) and 20 songs each from five female singers in the Korean ballad style (BS).
All recordings were sampled at 48kHz with 16-bit quantization.
For each singer, the songs were separated into two parts (90\% and 10\%) for training and testing, respectively.
All songs were segmented into a range of 5 to 15 seconds.
For the acoustic features, the 80-dimensional log mel-spectrograms were extracted with a fast Fourier transform size, window size, and frame shift of 2,048, 1,920, and 480, respectively.
Further, the 60-dimensional MGC, 5-dimensional BAP, F0, and V/UV flag were extracted by WORLD with a hop size of 10 ms. 
We used additional data from the open source datasets Kiritan\cite{kiritan}, CSD\cite{csd}, and VocalSet\cite{vocalset} to improve the quality of the vocoder.

\subsubsection{Model configurations}
In our experiments, the phoneme, note, and singer ID were embedded as a 384-dimensional vector. 
All encoders and decoders were stacked with six conformer blocks\cite{conformer}, each of which were composed of the following modules: multi-head self-attention (MHSA), convolution, and feedforward (two).
In the MHSA module, the number of heads and the hidden size of self-attention were 2 and 384, respectively.
In the convolution module, the hidden size of pointwise convolutions was 384, and the kernel size of 1D depth wise convolutions was 31.
In the feedforward modules, the input/output sizes of the first and second linear layers were 384/1536 and 1536/384, respectively.
The dropout rate for each module of the conformer block was 0.1.
The FC layers of the mel-predictor described in Figure~\ref{fig:overview}(c) converted the two 384-dimensional hidden representations into timbre and pitch representations of 80 channels. 
The postnet architecture used is the same as the postnet in \cite{tacotron2}, where the kernel size and channels are set to 5 and 80, respectively, and all layers except the last layer use the tanh activation function.
The discriminators described in Figure~\ref{fig:overview}(d) consist of 1D convolution layers with leaky ReLU activation functions. 
For the unconditional layers of the discriminators, the number of channels, kernel sizes and strides of 1D convolution layers are [128,256,1024,256,1], [5,9,9,9,5], and [1,2,2,1,1].
For the conditional layers of the discriminators, the parameters of the conditional layers are the same as the 4th and 5th unconditional layers.
For the loss weights, we set $w_m$, $w_b$, $w_v$, $w_p$, and $\lambda_{fm}$ to 10, 1, 1, 4, and 10, respectively.
For the vocoder, we use Parallel WaveGAN (PWG) \cite{pwg}.
To reconstruct the singing voice waveform with a 48k sampling rate, we set the kernel size of each 1D-convolutional layer to 31 to have a large receptive field, while the rest are the same as \cite{hifisinger}.

\subsubsection{Training}
In the acoustic model, we first pre-trained for 30,000 iterations to disentangle the representations of timbre and pitch.
Then, we trained 270,000 iterations for the multi-task training. 
In all training phase, the generator is adversarial trained with the discriminators.
We used the Adam optimizer ($\beta_1=0.9$, $\beta_2=0.98$ and $\epsilon=10^{-9}$) with a batch size of 4.
The initial learning rate was set to 0.0001 and halved after every 50,000 iterations. 
All parameters of the generator and discriminator were initialized via the Xavier initialization\cite{xavier}.
We also trained the vocoder for 1,000,000 iterations using a RAdam\cite{RAdam} optimizer with the same hyperparameters and training method as in \cite{hifisinger}. 

\subsection{Results}
\subsubsection{Listening test}
We conducted the mean opinion scores (MOS) listening test on the overall performance including perceptual quality and naturalness. 
The proposed multi-task SVS model was compared with the single-task SVS model depicted in Figure~\ref{fig:overview}(a) and the N-Singer model.
The N-Singer was extended to a multi-singer model by adding singer embeddings to the two encoder outputs.
Additionally, the performance of the sample synthesizing the predicted WORLD vocoder features of the proposed model was also compared with XiaoiceSing, a conventional WORLD vocoder-based SVS model.
Moreover, the XiaoiceSing model was extended to a multi-singer model in the same way as above, and was adversarial trained for logF0 and MGC using the discriminators of the proposed model.
We included the reconstructed samples (as the ground truth) to check the upper bound of the vocoder's performance.
In each model, 32 samples were prepared by randomly selecting 8 samples each for four singers (two CS singers and two BS singers).
Then, we requested 13 native-speaking Korean participants to evaluate the MOS for the overall performance. 

\begin{table}[t]
  \caption{The MOS test results with 95\% confidence intervals}
  \label{tab:mos1}
  \centering
  \renewcommand{\arraystretch}{1}
  \renewcommand{\tabcolsep}{3.5mm}
  \begin{tabular}{l|c|c}
    \toprule
    Method & PWG  & WORLD\\
    \midrule
        \textit{XiaoiceSing}              & - & 3.35 $\pm$ 0.09  \\
        \textit{N-Singer}                 & 3.53 $\pm$ 0.09 & -  \\
        \textit{Single-task}              & 3.35 $\pm$ 0.08 & -  \\
        \textit{Multi-task (Proposed)}    & \textbf{3.76} $\pm$ \textbf{0.09} & \textbf{3.75} $\pm$ \textbf{0.08}  \\
    \midrule
        \textit{Ground truth}             & 3.91 $\pm$ 0.10 & 4.23 $\pm$ 0.08  \\
    \bottomrule
  \end{tabular}
  \vspace{-0.4cm}
\end{table}

In Table\ref{tab:mos1}, for the PWG, the MOS results demonstrate that the proposed model performs better than the single-task model and N-Singer.
For the WORLD vocoder, the singing voices synthesized from the multi-task model exhibit better performance than the singing voices of the XiaoiceSing. 
It shows that the two decoders are more advantageous than the integrated decoder when modeling timbre and pitch. 
Although the performance of WORLD vocoder is better than that of PWG for ground truth, the proposed model is better for samples synthesized using the PWG than that of the WORLD.
In future work, improving the performance of the 48k sampling rate-based-vocoder for singing voices is left.

\subsubsection{Effect of adversarial training for auxiliary task}

\begin{figure}[t]
  \centering
  \includegraphics[width=\linewidth]{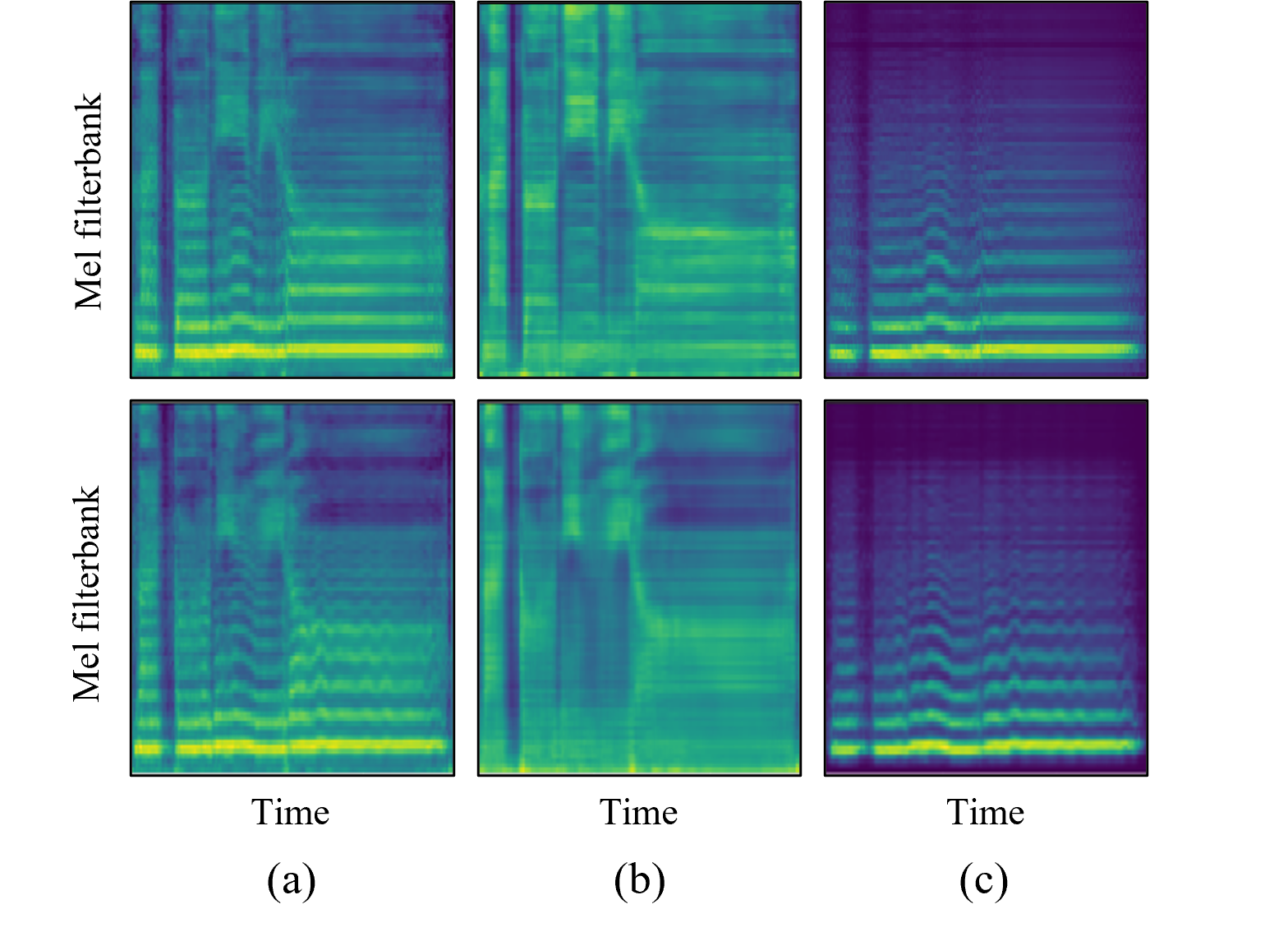}
  \caption{Generated mel-spectrograms of (a) final predicted output, (b) timbre representation, and (c) pitch representation. The top-line figures represent spectrograms from the model trained using adversarial loss for mel-spectrogram alone, while the bottom is from the model using adversarial loss for all features.}
  \label{fig:melspec}
   \vspace{-0.3cm}
\end{figure}

We conducted an ablation study to confirm the effect of the adversarial training on the auxiliary task.
Figure~\ref{fig:melspec} presents an example of a mel-spectrogram comparing two kinds of adversarial training (one with and another without the auxiliary task), where (a) is the synthesized final mel-spectrogram; (b) and (c) are the timbre and pitch representations, respectively.
The top line represents adversarial training applied to the mel-spectrogram alone, while the bottom is the adversarially trained mel-spectrogram with MGC and logF0.
The bottom panels of (b) and (c) show that the spectral envelope and pitch harmonics are separated.
In contrast, the top panels of (b) and (c) show that the pitch harmonic component remains in the timbre representations, while the pitch representation has a weak vibrato compared to the bottom.
Therefore, it shows that the precise predictions of MGC and logF0, which are the auxiliary tasks, help to disentangle the timbre and pitch representations.

\begin{figure}[t]
  \centering
  \includegraphics[width=\linewidth]{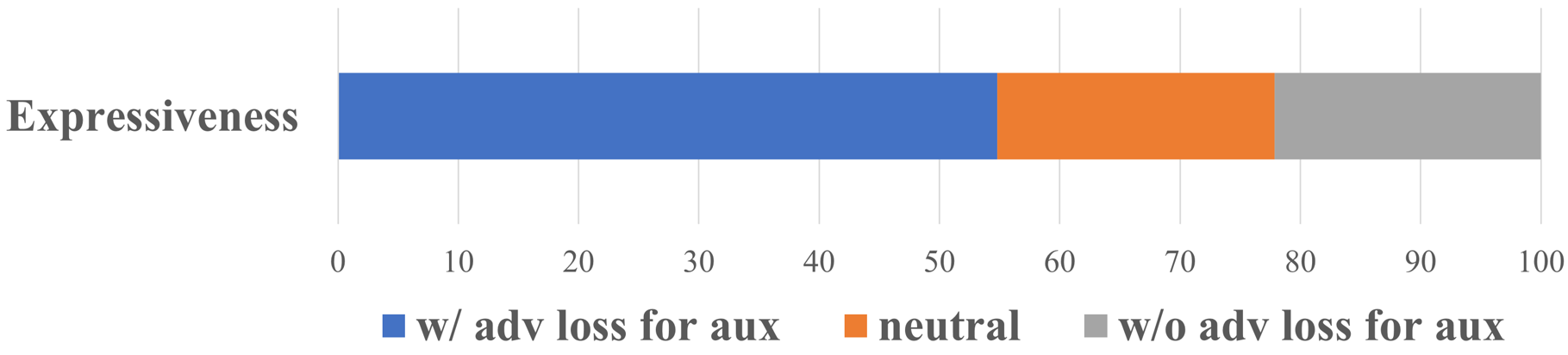}
  \caption{A/B preference test for expressiveness of the generated samples from the proposed model with or without adversarial training for auxiliary task.}
  \label{fig:abtest}
  \vspace{-0.6cm}
\end{figure}

We also conducted the A/B preference test to compare the performance difference according to the disentangling ability.
For the A/B test, 16 samples of singing voices generated for two BS singers were prepared. 
Then, we asked the participants to evaluate the expressiveness of the pitch curve and energy dynamics.
As shown in Figure~\ref{fig:abtest}, 54.8\% preferred the proposed model using adversarial loss for the auxiliary task, while 22.1\% did not.
The remaining 23.1\% were neutral.
A chi-squared test on the result classes had a p-value of $1.46 \times 10^{-11}$, implying that the A/B test result was significant.
Therefore, these results confirm that the accurate prediction of MGC and logF0 through the auxiliary tasks helps the decoder better represent the timbre and pitch of singing voices. 
\vspace{-0.1cm}

\section{Conclusions}
In this study, we propose an adversarial MTL-based SVS to disentangle the timbre and pitch representations.
The proposed approach demonstrates the ability to disentangle timbre and pitch by considering the interdependencies between them.
Experimental results confirm that the proposed model performs better than single-task SVS models.
Further, the samples synthesized by the auxiliary features also exhibited better performance than that of the conventional WORLD vocoder-based SVS.
Some audio samples are available online\footnote{https://nc-ai.github.io/speech/publications/amtl-svs/}.

\bibliographystyle{IEEEtran}

\bibliography{main}

\end{document}